# Interplay between magnetism and superconductivity in $EuFe_{2-x}Co_xAs_2$ studied by $^{57}Fe$ and $^{151}Eu$ Mössbauer spectroscopy

A. Błachowski [1], K. Ruebenbauer [1*], J. Żukrowski [2], Z. Bukowski [3,4], K. Rogacki [3], P. J. W. Moll [4], and J. Karpinski [4]

[1] Mössbauer Spectroscopy Division, Institute of Physics, Pedagogical University
PL-30-084 Kraków, ul. Podchorążych 2, Poland

[2] Solid State Physics Department, Faculty of Physics and Applied Computer Science, AGH University of Science and Technology
PL-30-059 Kraków, Al. Mickiewicza 30, Poland

[3] Institute of Low Temperatures and Structure Research, Polish Academy of Sciences
PL-50-422 Wrocław, ul. Okólna 2, Poland

[4] Laboratory for Solid State Physics, ETH Zurich
CH-8093 Zurich, Switzerland

[*] Corresponding author: sfrueben@cyf-kr.edu.pl

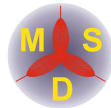

## Abstract

The compound $EuFe_{2-x}Co_xAs_2$ was investigated by means of the $^{57}Fe$ and $^{151}Eu$ Mössbauer spectroscopy versus temperature (4.2 – 300 K) for x=0 (parent), x=0.34 – 0.39 (superconductor) and x=0.58 (overdoped). It was found that spin density wave (SDW) is suppressed by Co-substitution, however it survives in the region of superconductivity, but iron spectra exhibit some non-magnetic component in the superconducting region. Europium orders anti-ferromagnetically regardless of the Co concentration with the spin re-orientation from the a-axis in the parent compound toward c-axis with the increasing replacement of iron by cobalt. The re-orientation takes place close to the a-c plane. Some trivalent europium appears in $EuFe_{2-x}Co_xAs_2$ versus substitution due to the chemical pressure induced by Co-atoms and it experiences some transferred hyperfine field from $Eu^{2+}$. Iron experiences some transferred field due to the europium ordering for substituted samples in the SDW and non-magnetic state both, while the transferred field is undetectable in the parent compound. Superconductivity coexists with the 4f-europium magnetic order within the same volume. It seems that superconductivity has some filamentary character in $EuFe_{2-x}Co_xAs_2$ and it is confined to the non-magnetic component seen by the iron Mössbauer spectroscopy.

## 1. Introduction

The EuFe$_2$As$_2$ is a parent metallic compound of the iron-based superconductors belonging to the '122' family. It crystallizes within tetragonal unit cell with a small orthorhombic distortion below ~190 K [1]. A transition to the orthorhombic phase has small hysteresis on the temperature scale, and it is accompanied by the development of the spin density wave (SDW) appearing just below transition. SDW has origin in the 3d band of iron and it is longitudinal wave having propagation direction and magnetic moment aligned with the crystallographic a-axis [2, 3]. The length of SDW is incommensurate with the lattice period along the a-axis. SDW is incoherent just below onset of the magnetic ordering and becomes coherent upon lowering temperature, i.e., at about 189 K SDW is already fully coherent [4]. Europium is located in the planes perpendicular to the c-axis separating [Fe$_2$As$_2$] layers. It stays in the divalent $^8S_{7/2}$ state without orbital contribution to the 4f magnetic moment and orders magnetically at about 19 K with magnetic moments aligned along the a-axis. Subsequent europium bearing planes are ordered in the anti-ferromagnetic fashion. Europium and iron nuclei experience almost axially electric field gradient (EFG) with the principal component aligned with the c-axis [2, 3].

Superconductivity could be achieved in EuFe$_2$As$_2$ either applying pressure [5-7] or by partial substitution e.g. either europium by potassium [8], arsenic by phosphorus [9], or iron by cobalt [10, 11]. One can obtain underdoped non-superconducting material, superconductor of the second type, and finally overdoped material without superconductivity, while increasing dopant concentration. All these compounds exhibit metallic behavior. SDW order becomes weaker with the increasing concentration of the dopant, and finally it vanishes in the overdoped region. The europium magnetic ordering temperature is very weakly perturbed by the dopant concentration [9] as long as dopants do not substitute europium itself [8].

This contribution is concerned with the EuFe$_{2-x}$Co$_x$As$_2$ compound investigations by means of the $^{57}$Fe and $^{151}$Eu Mössbauer spectroscopy versus temperature and cobalt concentration x.

## 2. Experimental

Single crystals of EuFe$_{2-x}$Co$_x$As$_2$ were grown applying tin flux method as described in Refs [11, 12]. They appeared as single-phase material according to the X-ray diffraction results. The cobalt concentration x was determined by using EDX analysis. Relative error is estimated as about 5 % and some overestimation could be expected due to the proximity of the cobalt, iron and europium fluorescent X-ray lines. Resistivity measurements have been performed versus temperature on single crystals in a four-point configuration for all dopant concentrations including parent compound and in the null external magnetic field. Results are reported here as relative resistivity, i.e., normalized to the resistivity at 300 K for each sample. Magnetic susceptibility versus temperature and magnetization versus external field for several temperatures was measured for x=0.37 single crystal.

Mössbauer absorbers for the $^{57}$Fe spectroscopy were prepared in the powder form in the same manner as described in Ref. [4]. Absorbers for $^{151}$Eu spectroscopy were made in the same way, albeit some of them contained about twice as much material per unit area. $^{57}$Fe spectra were collected using the same equipment and procedures as described in Ref. [4]. $^{151}$Eu spectra were collected applying $^{151}$SmF$_3$ source kept at room temperature and a scintillation detector. Spectra were processed within the transmission integral approximation by using applications from the MOSGRAF-2009 suite [13]. Iron spectra with SDW component were processed by GMFPHARM application treating the electric quadrupole interaction in the first order approximation. Remaining spectra were processed by GMFP application and the full Hamiltonian was diagonalized in both nuclear states. The europium hyperfine anomaly was accounted for. Spectral shifts are reported versus room temperature α-Fe or versus room temperature $^{151}SmF_3$ source, respectively.

## 3. Results

The temperature evolution of the resistivity is given in Figure 1 for various cobalt concentrations x. Upon entering the SDW state, the metallic behavior could be changed significantly due to the partial gapping of the Fermi surface leading to the decrease of the carrier concentration. On the other hand, the spin scattering is reduced owing to the increasing spin order. The first mechanism leads to the upturn, while the second to the downturn of the resistivity with lowering of the temperature. A reduced carrier concentration is clearly seen for x=0.34 sample. Additionally, the divalent europium magnetic ordering is seen as much less pronounced kink on the resistivity due to further reduction of the spin scattering. This kink could be seen for the parent compound and for the overdoped x=0.58 sample, as otherwise it is masked by the much stronger effect due to the development of the superconductivity. The effect of the europium ordering on the resistivity is much lesser for the overdoped material in comparison with the parent compound. Hence, one can conclude that the coupling between 4f and conduction electrons is weak. The zero resistance in the superconducting state has been observed only for x=0.39 sample.

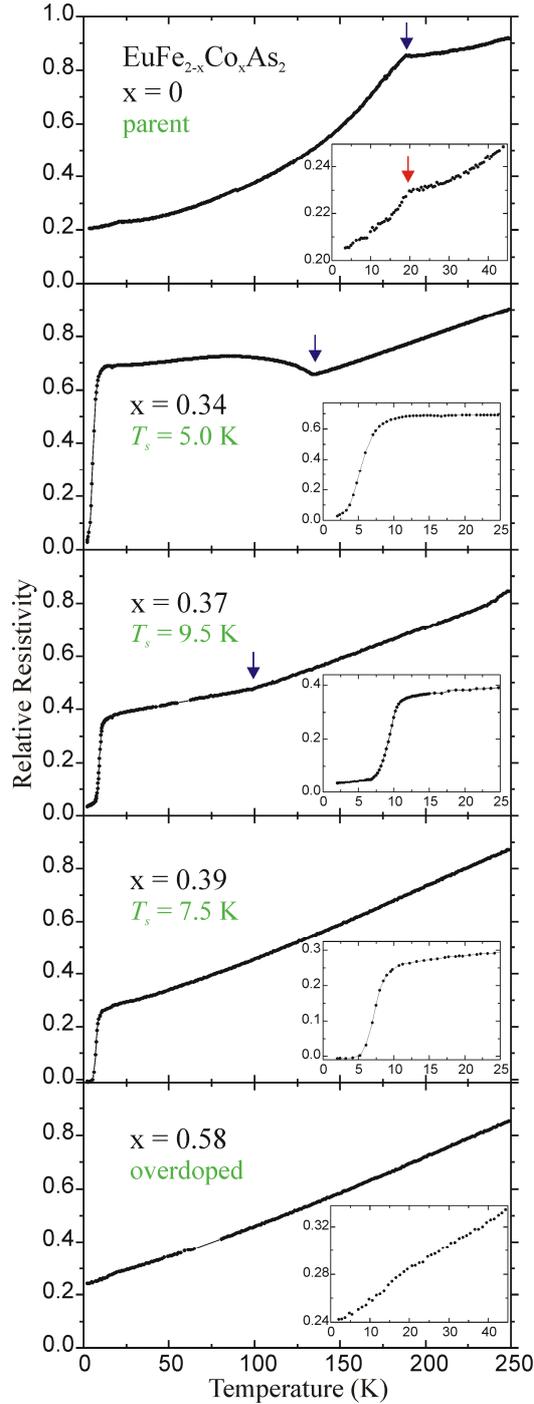

**Figure 1** Relative resistivity (normalized to the resistivity at 300 K for each sample) plotted versus temperature for all samples investigated. The current was applied in the a-b plane. Blue arrows indicate change of the slope due to the SDW development. Red arrow shows change of the slope due to the europium magnetic ordering in the parent compound. Note that zero resistance is obtained only for x = 0.39 sample. The significant drop of the resistivity is observed for x = 0.34 and x = 0.37 samples due to development of the superconductivity. $T_s$ denotes transition temperature to the superconducting state. Insets show expanded regions of the low temperatures.

Results of the magnetic measurements are shown in Figure 2. A magnetic susceptibility $\chi'_{ac}$ shows some small diamagnetic deviation at 10 kOe and below 5 K for x=0.37 sample indicating that part of the sample is in the superconducting state. This feature is particularly pronounced for the zero-field-cooled (ZFC) state of the material. A significant hump observed for the susceptibility (x=0.37) measured in the zero field (see, Figure 2a) is due to the magnetic ordering of divalent europium. The presence of such hump is an indication that the anti-ferromagnetic order of the europium atoms (observed in the parent compound) is perturbed by cobalt replacing iron [14]. The ac magnetic susceptibility data obtained for x=0.37 sample at 200 K yield the effective magnetic moment $\mu_{eff} = 7.94\,\mu_B$ ($\mu_B$ stands for the Bohr magneton) in good agreement with the value obtained from the saturation magnetization at low temperature (see, Figure 2b). Such an effective moment is due to the divalent europium with $\mu_{eff} = g\mu_B\sqrt{S(S+1)}$, where the atomic giro-magnetic factor amounts to $g \approx 2$ and the respective spin of the atomic shell equals $S = 7/2$. The saturation magnetization at low temperature (see, Figure 2b) yields magnetic moment $6.2\,\mu_B$ in the ordered state being in fair agreement with the susceptibility data. Results of the electric and magnetic measurements strongly suggest filamentary character of the superconductivity.

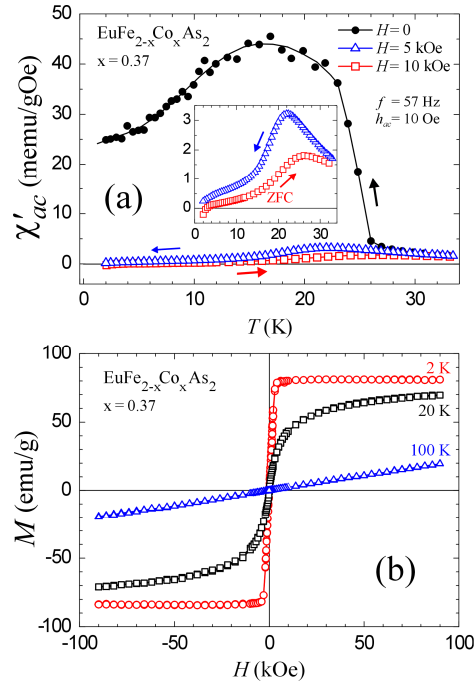

**Figure 2** Section (a) shows real part of the magnetic susceptibility $\chi'_{ac}$ plotted versus temperature $T$ for the crystal with composition $x = 0.37$. A dc field of $H = 0$, 5 and 10 kOe was applied along the c-axis. The ac field of $h_{ac} = 10$ Oe and frequency $f = 57$ Hz was applied in the same direction. Results were obtained for zero-field-cooled (ZFC) and field-cooled states. The inset shows expanded vertical scale for $H = 5$ and 10 kOe. Diamagnetic behavior below 5 K is more pronounced for $H = 10$ kOe and it was obtained in the ZFC state. Section (b) shows magnetization $M$ loops obtained at 2, 20 and 100 K versus external field $H$ for the sample with composition $x = 0.37$. The field $H$ was applied along the c-axis. For the lowest temperature of 2 K magnetization $M$ saturates in the field of about 3 kOe at the value of 83 emu/g corresponding to the magnetic moment in the ordered state of 6.2 Bohr magnetons per chemical formula.

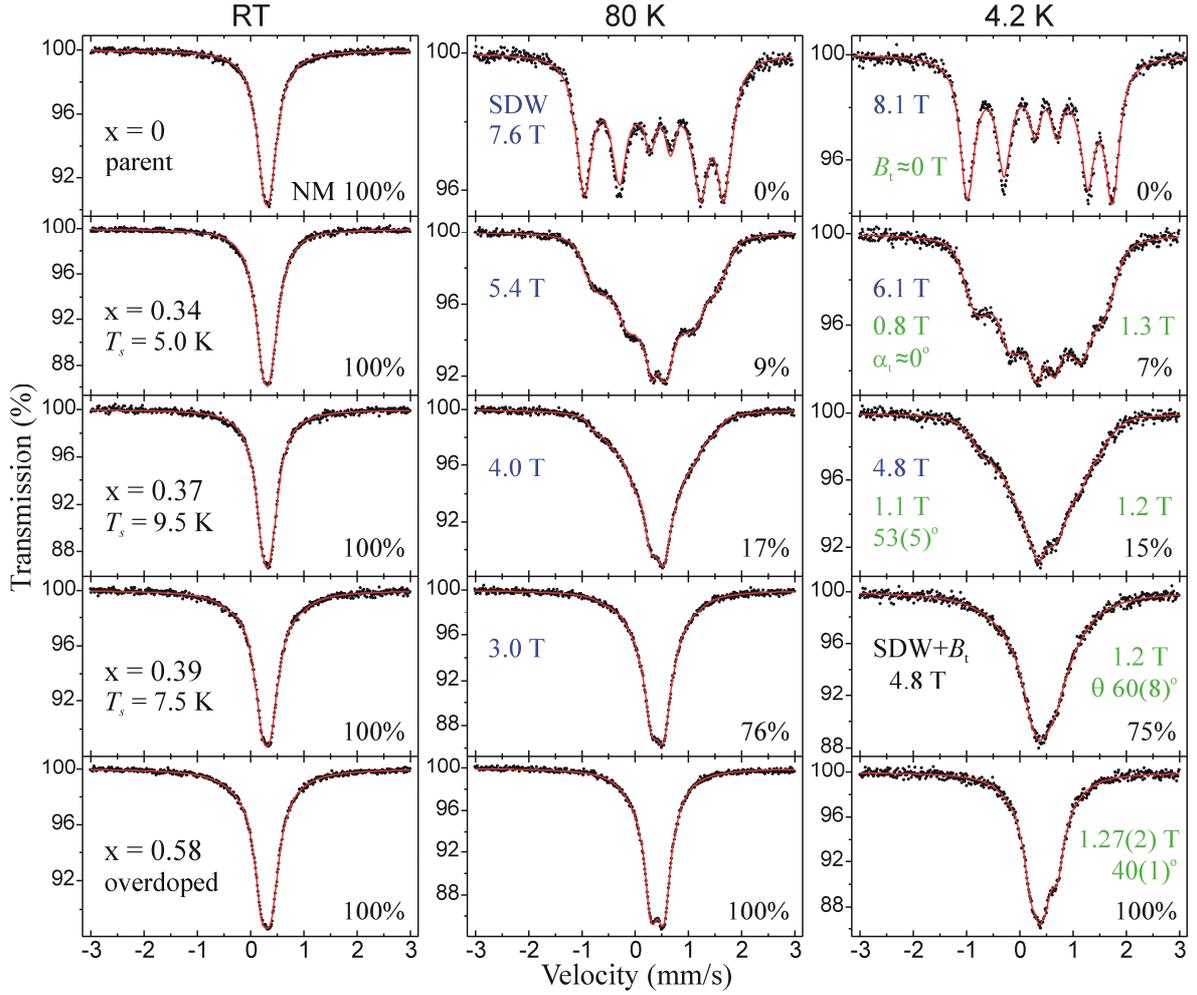

**Figure 3** $^{57}$Fe Mössbauer spectra of EuFe$_{2-x}$Co$_x$As$_2$ for various concentrations x obtained at room temperature (RT), 80 K and 4.2 K. Contribution due to the non-magnetic (NM) component is shown in the lower right corner of each spectrum. Amplitudes $\sqrt{\langle B^2 \rangle}$ of the SDW fields are shown in blue. Transferred field on iron for the SDW component is shown in green on the left together with the angle $\alpha_t$. For x=0.39 it is impossible to separate SDW and transferred fields for the SDW component. Transferred field on iron for the NM component is shown in green on the right together with the angle $\theta$.

Figure 3 shows selected $^{57}$Fe spectra obtained versus temperature and cobalt concentration x. Spectra were fitted with the SDW model [4] in the magnetically ordered region and with a quadrupole doublet otherwise. The shape of SDW and corresponding hyperfine magnetic field distributions are shown in Figure 4. For all spectra obtained at 4.2 K, i.e., below onset of the europium magnetic order one has to take into account transferred hyperfine field on iron. Essential results are summarized in Table I and in Figure 3. For x=0 (parent compound) results have been already published [4]. There is no non-magnetic component below transition to the SDW-state. SDW approaches *quasi*-rectangular shape close to saturation. The transferred field on iron due to the magnetic ordering of europium is undetectable.

**Table I**

Selected results obtained from $^{57}$Fe Mössbauer spectra shown in Figure 3. Contributions *A* to the spectral shape due to the SDW component and non-magnetic (NM) components are listed with the corresponding spectral shift S versus room temperature α-Fe and quadrupole splitting Δ. For spectra without SDW component $\Delta = \frac{1}{2}(c/E_0)eQ_eV_{zz}$, while for spectra with the SDW component $\Delta = \frac{1}{4}(c/E_0)eQ_eV_{zz}(3\cos^2\theta - 1)$ - see Ref. [4]. Absorber line widths are 0.17 mm/s for SDW components, 0.2 mm/s for NM or NM1 components and 0.5 mm/s for NM2 components. Errors for all values are of the order of unity for the last digit shown.

| | | A (%) | S (mm/s) | Δ (mm/s) |
|---|---|---|---|---|
| x = 0 | | | | |
| RT | NM | 100 | 0.425 | 0.118 |
| 80 K | SDW | 100 | 0.540 | -0.120 |
| 4.2 K | SDW | 100 | 0.552 | -0.116 |
| x = 0.34 | | | | |
| RT | NM | 100 | 0.422 | 0.132 |
| 80 K | SDW | 91 | 0.527 | -0.079 |
| | NM | 9 | 0.53 | 0.18 |
| 4.2 K | SDW | 93 | 0.539 | -0.097 |
| | NM | 7 | 0.54 | 0.38 |
| x = 0.37 | | | | |
| RT | NM | 100 | 0.422 | 0.130 |
| 80 K | SDW | 83 | 0.533 | -0.09 |
| | NM | 17 | 0.53 | 0.24 |
| 4.2 K | SDW | 85 | 0.54 | -0.09 |
| | NM | 15 | 0.54 | 0.33 |
| x = 0.39 | | | | |
| RT | NM1 | 88 | 0.415 | 0.165 |
| | NM2 | 12 | 0.38 | 1.2 |
| 80 K | SDW | 24 | 0.54 | -0.24 |
| | NM1 | 68 | 0.54 | 0.23 |
| | NM2 | 8 | 0.46 | 0.8 |
| 4.2 K | SDW | 25 | 0.55 | -0.2 |
| | NM1 | 67 | 0.55 | 0.2 |
| | NM2 | 8 | 0.55 | 0.9 |
| x = 0.58 | | | | |
| RT | NM1 | 81 | 0.415 | 0.180 |
| | NM2 | 19 | 0.32 | 0.78 |
| 80 K | NM1 | 81 | 0.536 | 0.234 |
| | NM2 | 19 | 0.43 | 0.63 |
| 4.2 K | NM1 | 81 | 0.548 | 0.20 |
| | NM2 | 19 | 0.53 | 0.92 |

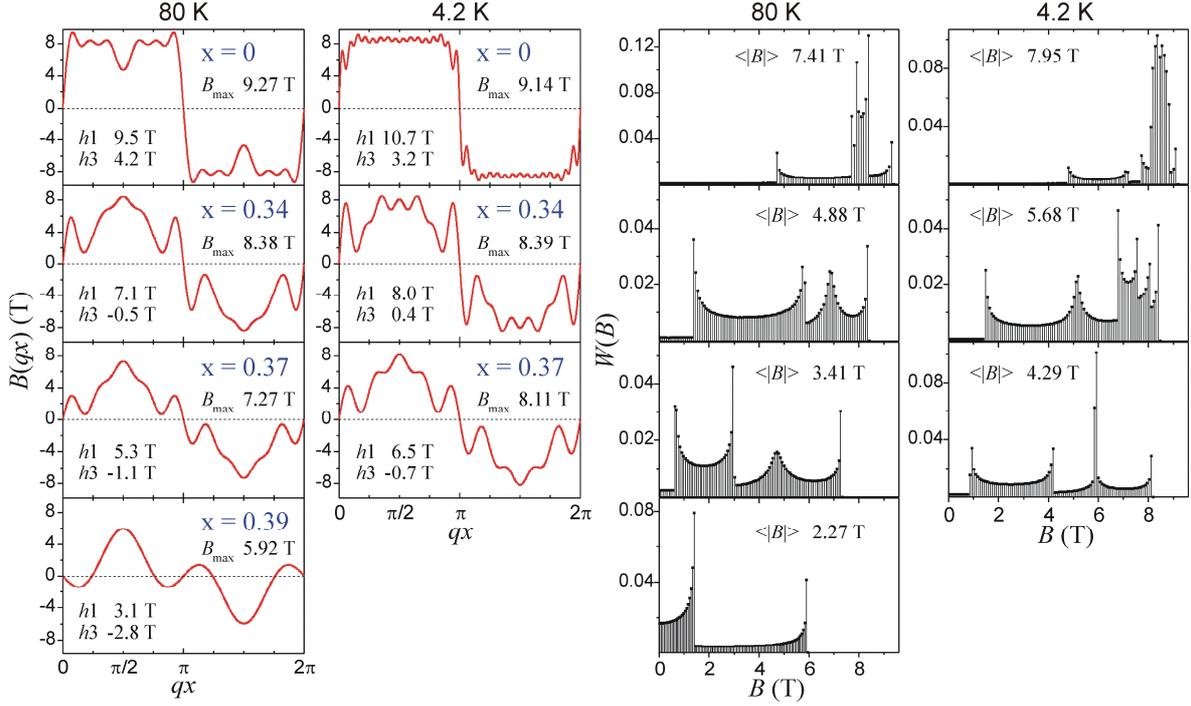

**Figure 4** Shape of SDW and corresponding hyperfine field distribution of $EuFe_{2-x}Co_xAs_2$ obtained at 80 K and 4.2 K for various Co concentrations x. Amplitudes of the first two harmonics $h_1$ and $h_3$ are shown. The maximum amplitude of SDW $B_{max}$ is shown, too. The symbol $qx$ denotes phase of SDW. The symbol $\langle|B|\rangle$ stands for the average field of the distribution.

Some non-magnetic component appears for superconductors with $x = 0.34 - 0.39$. The non-magnetic contribution to the spectrum increases with the increasing cobalt concentration, and it survives even at 4.2 K. The amplitude of SDW in the ground state diminishes with increasing x. The shape of SDW evolves with the increase of the Co concentration in similarity to the shape evolution with the increasing temperature [4]. The evolution versus dopant concentration exhibits more erratic character than the evolution caused by the variation of the temperature. The irregularity is likely to be caused by fluctuations of the cobalt concentration. A transferred field appears in all substituted samples at 4.2 K. It is seen in the SDW and non-magnetic components. In the case of SDW component one can determine the angle $\alpha_t$ between SDW field and the transferred field (see, Figure 5 for the geometry of hyperfine fields and electric field gradient tensor). For the x=0.34 sample one obtains $\alpha_t \approx 0$, i.e., the transferred field is oriented along the a-axis parallel/anti-parallel to the SDW field. For x=0.37 transferred field makes an angle $\alpha_t \approx 53°$ with the SDW field, i.e., a-axis. For x=0.39 one cannot separate hyperfine field into SDW and transferred components. For the non-magnetic component one can determine the angle θ between transferred field and the principal EFG component aligned along the c-axis starting since x=0.39. For x=0.58 (overdoped) one has sole non-magnetic component with the transferred field below europium magnetic ordering temperature. Broadening of the 4.2 K spectrum of this sample is due to the transferred field, as SDW does not develop for such high concentration of cobalt. Hence, the angle $θ = 40°$ is determined most reliably from this spectrum. The non-magnetic component could be described by single doublet till x=0.37. For higher cobalt concentrations two doublets are needed to describe spectral shape with different quadrupole splitting and slightly different spectral shift [9]. The quadrupole interaction is treated in the first order approximation for spectra containing SDW components and the full Hamiltonian is used for the spectra with the transferred field alone or with the negligible contribution from SDW to the total magnetic field. The coupling constant in the first order approximation is proportional to $V_{zz}(3\cos^2\theta - 1)$ with the symbol $V_{zz}$ denoting principal component of the EFG tensor. SDW spectra yield above term negative with the angle $\theta$ being the right angle due to the local symmetry. Hence, the principal component of the EFG tensor is positive in these compounds. Note that the quadrupole interaction appears solely for the excited nuclear state in the case of the resonant transition in iron, and that the spectral quadrupole moment of this state is positive. On the other hand, the component $V_{zz}$ has been found as negative in FeSe [15]. It is interesting to note, that the quadrupole splitting (proportional to $V_{zz}$) at room temperature varies for the parent compounds of the '122' family from almost zero for $BaFe_2As_2$ to 0.194 mm/s for $CaFe_2As_2$ [4]. Hence, the charge distribution around iron is perturbed by the atomic layers with the A-type atoms (A=Ca, Sr, Ba, Eu). However, the local symmetry around iron is preserved, as the $[Fe_2As_2]$ layers are almost the same in all of these compounds.

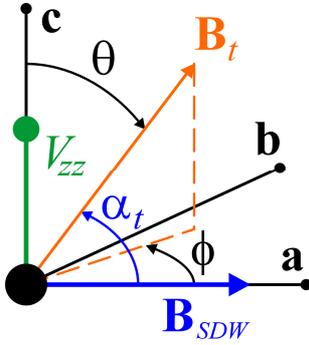

**Figure 5** Geometry of the SDW field $\mathbf{B}_{SDW}$, transferred field $\mathbf{B}_t$ and the principal component of the axially symmetric EFG tensor $V_{zz}$ in the crystal axes **abc** of the orthorhombic unit cell as seen by the iron nucleus. Note that for $\phi = 0$ one obtains $\alpha_t + \theta = 90°$.

Figure 6 shows $^{57}$Fe spectra of the sample x=0.34 versus temperature. SDW component appears at approximately 160 K with about 70 % contribution to the spectrum area. The SDW component contribution amounts to 93 % at 4.2 K. The SDW shape for the sample with x=0.34 could be fitted with five subsequent odd harmonics in the whole temperature range. The spectral shift is practically the same for SDW and non-magnetic components indicating that the electron density on the iron and atomic dynamics of the iron atoms are the same. On the other hand, the electron spin order is different, as it is extremely sensitive to the concentration of the dopant and it sees tiny variations of the cobalt concentration. The square root from the mean squared amplitude of the SDW $\sqrt{\langle B^2 \rangle}$ is plotted versus temperature in Figure 7 for the parent compound, and for x=0.34 and x=0.37 samples. Data were treated in the same way as in Ref. [4], and the meaning of the symbols is the same. Table II gathers essential results for x=0.34 sample. Results obtained previously for the parent compound are repeated for comparison [4]. It was found that for x=0.34 coherent SDW vanishes above temperature $T_c = 150\,\text{K}$, while the incoherent SDW vanishes below temperature $T_0 = 140\,\text{K}$. A difference between these two temperatures is more than three times larger than similar difference in the parent compound. Such effect is understandable taking into account fluctuations of the dopant concentration across the sample. The critical exponent in the coherent region $\alpha_0$ is the same as in the parent compound. The universality class remains (1, 2) after substitution indicating that the electronic spin system of SDW obeys Ising model (one dimensional spin space) and it has two dimensions in the configuration space (magnetized planes). The parameter $\gamma$ is equal zero showing that the system remains critical in the whole temperature range between $T_c$ and the ground state. The exponent $\beta$ in the incoherent region decreased in comparison with the parent compound due to the inhomogeneities introduced by the dopant fluctuations, and therefore incoherent region covers wider span of temperature than in the parent.

**Table II**

Essential parameters describing evolution of $\sqrt{\langle B^2 \rangle}$ versus temperature. Symbols have the following meaning: $B_0$ - saturation field, $B_F$ - field at bifurcation into coherent and incoherent parts. The symbol $T_c$ is the temperature at which coherent part appears upon cooling. The symbol $T_0$ stands for the temperature at which incoherent part appears upon heating, $\alpha_0$ is a critical exponent below transition, $\gamma$ stands for the parameter describing evolution of the exponent upon cooling to the ground state, and $\beta$ denotes exponent describing evolution of the incoherent part. For details see Ref. [4].

| Co-concentration | x = 0 | x = 0.34 |
|---|---|---|
| $B_0$ (T) | 8.03(2) | 5.99(4) |
| $B_F$ (T) | 4.8(1) | 4.3(3) |
| $T_c$ (K) | 192.1(1) | 150(2) |
| $T_0$ (K) | 189.1(1) | 140(2) |
| $\alpha_0$ | 0.124(1) | 0.125(9) |
| $\gamma$ | 0.9(1) | 0 |
| $\beta$ | 7.6(1) | 1.8(1) |

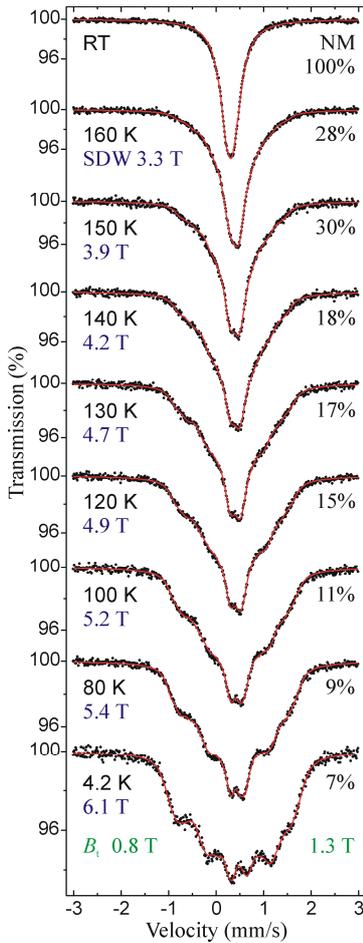

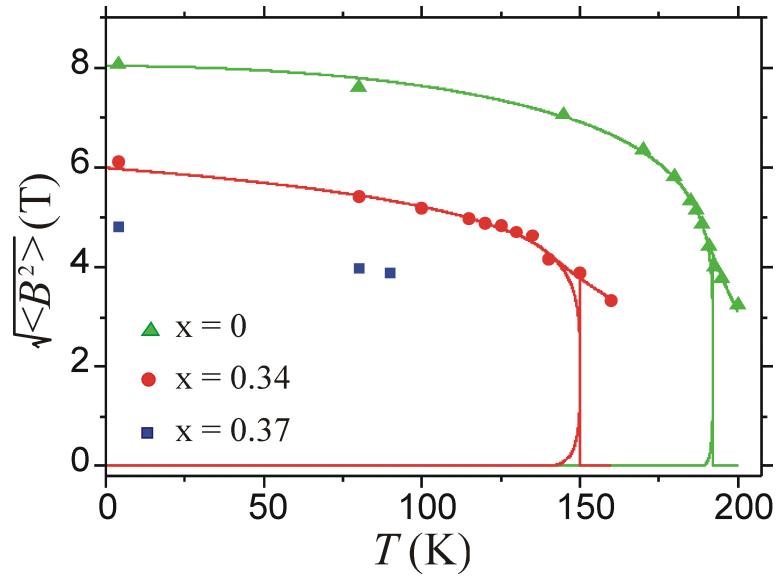

**Figure 6** Selected $^{57}$Fe Mössbauer spectra obtained versus temperature for x=0.34. Amplitudes $\sqrt{\langle B^2 \rangle}$ of the SDW fields are shown in blue. A contribution due to the non-magnetic (NM) component is shown. Transferred field $B_t$ on iron is shown in the left for SDW component and in the right for the NM component of the spectrum.

**Figure 7** Plot of the SDW amplitude $\sqrt{\langle B^2 \rangle}$ versus temperature for various Co concentrations x. Solid lines represent total, coherent and incoherent contributions as described in Ref. [4]. For x=0.37 there is not enough spectral SDW component and the SDW amplitude is too small to obtain reliable results.

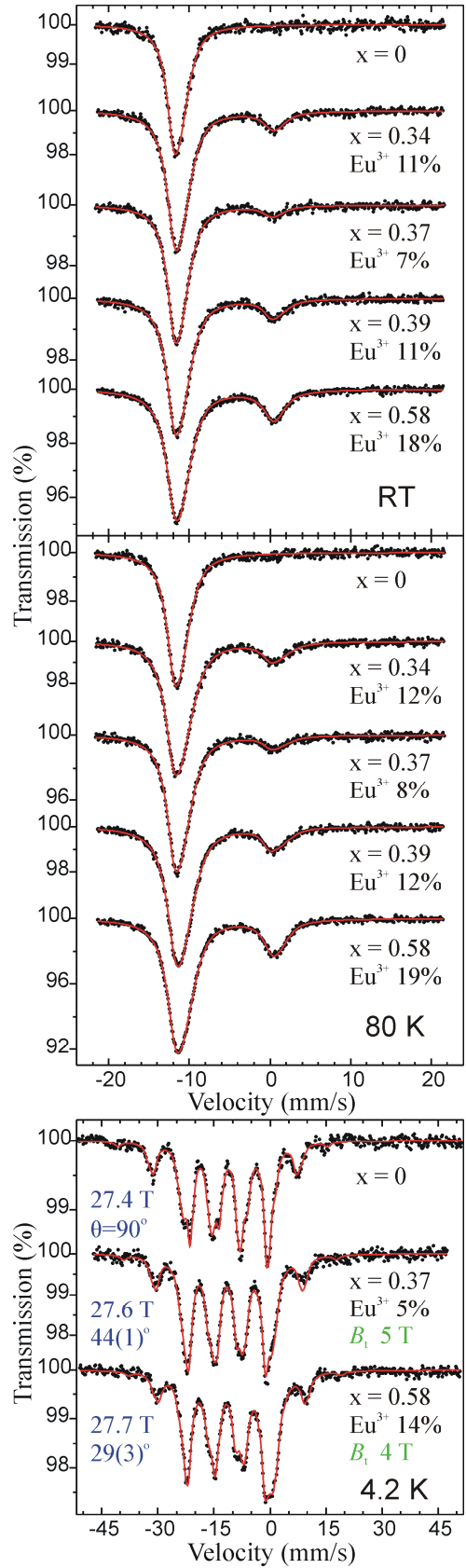

**Figure 8** $^{151}$Eu Mössbauer spectra of EuFe$_{2-x}$Co$_x$As$_2$ for various Co concentrations x obtained at RT, 80 K and 4.2 K. Apparent contribution due to Eu$^{3+}$ is shown. Hyperfine magnetic field and the angle θ for Eu$^{2+}$ are shown in blue. Transferred field on Eu$^{3+}$ is shown in green.

Figure 8 shows $^{151}$Eu Mössbauer spectra. Europium occurs in the divalent state for the parent compound solely. The spectral shift varies from $-11.5$ mm/s at RT till $-11.3$ mm/s at 4.2 K. For Co-substituted samples corresponding spectral shift ranges between $-11.4$ mm/s and $-11.2$ mm/s. Some trivalent europium appears upon substitution with the spectral shift 0.5 mm/s at RT and 0.7 mm/s at 4.2 K. The amount of the trivalent europium increases with the increasing cobalt concentration. The EFG tensor is expected to be almost axially symmetric in both phases, i.e., tetragonal and orthorhombic with the principal component aligned along the c-axis. For both valency states, i.e., $^8S_{7/2}$ for divalent europium and $^7F_0$ for non-magnetic trivalent state there is no atomic contribution to the EFG as both states are spherically symmetric. Hence, the EFG is small and of the lattice origin. Note that the quadrupole moment $Q_g$ for the ground state of $^{151}$Eu is positive. It was found that the principal component of the EFG tensor is negative and very weakly depends on temperature and substitution. For parent compound the quadrupole coupling constant $\frac{1}{4}(c/E_0)eQ_gV_{zz}$ ranges between $-1.5$ mm/s at RT and $-1.7$ mm/s at 4.2 K. For Co-substituted samples there is not much difference as the respective range is between $-1.6$ mm/s and $-1.9$ mm/s. The quadrupole coupling constant is practically the same for both valency states indicating very similar local arrangement of atoms surrounding these sites. Close proximity of EFG acting on both Eu states is a strong hint that $Eu^{3+}$ is located in the same phase as $Eu^{2+}$. It has been reported that $Eu^{2+}$ transforms into $Eu^{3+}$ upon applying hydrostatic pressure to the parent compound [16-18], and applying local chemical pressure due to substitution in the EuFe$_2$As$_{1.4}$P$_{0.6}$ superconductor [17]. Generally one can expect transformation of any divalent europium ion into trivalent state provided applied pressure is sufficiently high. Such effect has been observed for pure metallic europium [19]. Here the similar effect occurs due to the local chemical pressure induced by cobalt substitution. Trivalent europium was previously observed in the substituted samples of Eu-122 compounds and interpreted as unidentified foreign phase [8, 9, 20]. The apparent contribution from $Eu^{3+}$ is the smallest at 4.2 K for a given sample. It means that trivalent europium is more strongly bound to the lattice than divalent europium and exhibits enhanced recoilless fraction.

$^{151}$Eu spectra obtained at 80 K do not show any magnetic broadening despite fully developed iron-based SDW at this temperature for parent and lightly substituted compounds. It means that europium nuclei are well shielded from the SDW field by the local electrons. Spectra obtained at 4.2 K exhibit hyperfine field typical for the magnetically ordered $Eu^{2+}$. A component due to $Eu^{3+}$ broadens significantly at this temperature due to the presence of the small magnetic field transferred from divalent europium. Hence, this is an additional argument in favor of the statement that $Eu^{3+}$ has to be located in the same phase as the dominant $Eu^{2+}$ ions. One has to bear in mind that trivalent europium has no magnetic moment. The angle $\theta$ between hyperfine field and the principal component of the EFG tensor (c-axis) is equal right angle for the parent compound. It means that europium hyperfine field (magnetic moment) is perpendicular to the c-axis in accordance with the established magnetic structure [2, 3]. The hyperfine field tilts on the c-axis with substitution. Europium spin re-orientation was observed previously in Eu-122 versus substitution with phosphorus [9]. It is impossible to fit independently angle $\theta$ for trivalent europium and it was assumed that it is the same as on divalent europium.

The sum of the angle $\alpha_t$ and the angle $\theta$ on europium is close to the right angle (for x=0.37: $\alpha_t + \theta = 53(5)^\circ + 44(1)^\circ$) indicating that europium spins re-orient close to the a-c plane – see Figure 5. A transferred field on iron follows approximately direction of the divalent europium magnetic moment [9, 21], as for x=0.58 angles θ amount to 40(1)° and 29(3)° for iron and europium, respectively. Magnetic susceptibility data (see, Figure 2a) suggest that for substituted samples the compensation of the europium moments to the anti-ferromagnetic state is incomplete. This result is in agreement with the Mössbauer results, but the helical order of the europium moments is rather unlikely due to the fact that they are oriented close to the a-c plane. Europium orders magnetically inside superconductor and inside SDW or overdoped material. Hence, one can expect that transfer field on iron has a character of the local dipolar field. However, the absence of such field in the parent compound shows that Eu-bearing planes are seen by iron rather like uniformly magnetized planes and not as individual atoms. Some local perturbation due to the presence of the dopant is essential to get transferred field on iron. Coexistence of the 4f magnetic order (predominantly anti-ferromagnetism here) and superconductivity is rare, but possible. The coexistence of the ferromagnetic order of the localized 4f magnetic moments of the $Gd^{3+}$ (iso-electronic with $Eu^{2+}$) and superconductivity of the second type was observed in the same volume for e.g. $Gd_xCe_{1-x}Ru_2$ cubic Laves phase [22]. Some kind of gadolinium magnetic order has been found in the superconducting $Gd_{0.84}Th_{0.16}FeAsO$ [23]. On the other hand, similar coexistence of the 3d magnetic order and superconductivity seems hardly possible for the Bose condensate composed of the Cooper pairs in the singlet state. Hence, it is more probable to have some filamentary superconductivity confined to the 3d non-magnetic regions of the sample. The bulk superconductivity is seen below the percolation limit of this filamentary structure [24].

## 4. Conclusions

Main results of the present contribution could be summarized as follows:

1. The SDW order diminishes in $EuFe_{2-x}Co_xAs_2$ with addition of cobalt i.e. a transition temperature is lowered together with the SDW amplitude. Substitution has similar effect on the SDW shape like temperature, albeit fluctuations are enhanced in comparison with the temperature effect.

2. SDW survives across the region of superconductivity and it vanishes in the overdoped region. However, in the region of superconductivity one has some non-magnetic component with the intensity increasing with the Co-substitution. It seems that superconductivity has some filamentary character.

3. Divalent europium orders magnetically regardless of the Co-substitution level. Europium moments rotate from the a-axis in the direction of the c-axis (within a-c plane) with increasing substitution. Some trivalent europium appears upon substitution with the amount increasing with the increasing cobalt concentration. $Eu^{3+}$ experiences some transferred field from $Eu^{2+}$. Europium magnetic order and superconductivity coexist in the same volume.

4. Iron experiences a transferred field from europium for the substituted material – in the SDW and non-magnetic state both. A transferred field is roughly parallel to the Eu magnetic moments.